\documentclass[12pt]{article}
\usepackage[dvips]{graphicx}

\def\bk{{\bf k}}
\def\bp{{\bf p}}
\def\br{{\bf r}}

\def\b\sigma{\mbox{\boldmath $\sigma$\unboldmath}}

\def\b\xi{\mbox{\boldmath $\xi$\unboldmath}}

\def\cl{{\cal L}}
\def\bX{{\bf X}}
\def\bR{{\bf R}}
\def\({\left(}
\def\){\right(}
\newcommand{\be}{\begin{equation}}
\newcommand{\ee}{\end{equation}}
\newcommand{\beq}{\begin{eqnarray}}
\newcommand{\eeq}{\end{eqnarray}}

\title{ \bf {Self-consistent equation for an interacting Bose gas}}
\author{Philippe A. Martin\\
Institute of Theoretical Physics\\ Swiss Federal Institute for Technology
Lausanne\\ CH-1015, Lausanne EPFL, Switzerland\\ \vspace{3mm}\\
Jaroslaw Piasecki\\
Institute of Theoretical Physics, University of Warsaw, Ho\.za 69 
\\00 681 Warsaw, Poland}

\date{\today}

\begin{document}

\maketitle

\begin{abstract} 
We consider interacting Bose gas in thermal equilibrium  
assuming a positive and bounded 
pair potential $V(r)$ such that $0<\int d\br V(r) = a<\infty$.
Expressing the partition function by the Feynman-Kac functional integral yields
a classical-like polymer representation of the quantum gas.
With Mayer graph summation techniques,
we demonstrate the existence of a self-consistent relation
$\rho (\mu)=F(\mu-a\rho(\mu))$ between the density $\rho $ and the chemical potential $\mu$,
valid in the range of convergence of Mayer series. 
The function $F$ is equal to the sum of all rooted multiply connected graphs. 
Using Kac's scaling $V_{\gamma}(\br)=\gamma^{3}V(\gamma r)$ we prove that in the 
mean-field limit $\gamma\to 0$ only tree diagrams contribute and function $F$ reduces 
to the free gas density. 

We also investigate how to extend the validity of the self-consistent relation beyond 
the convergence radius of Mayer series (vicinity of Bose-Einstein  condensation)
and study dominant corrections to mean field. At lowest order, the form of function $F$ 
is shown to depend on single polymer partition function for which we derive lower and upper 
bounds and on the resummation of ring diagrams which can be analytically performed.
\end{abstract}

\vskip 0.5cm 
{\bf KEYWORDS~:} Bose gas, quantum Mayer graphs, mean field, interacting polymers .
\vskip 0.5cm 
{\bf PACS numbers~:} {0530, 6700}
\vskip 0.5cm 
\noindent {\it Corresponding author ~:} Philippe A. Martin, \\
e-mail: phmartin@dpmail.epfl.ch

\section{Introduction}

The interest for a better understanding of Bose-Einstein condensation
has been strongly stimulated in recent years by the beautiful experimental
observations of condensates of cold atoms in traps [1,2].

Concerning rigorous results on the existence of Bose-Einstein condensation
in an interacting gas with pair interactions, we quote the work of Lieb and
Seiringer [3]. The authors show the existence of off-diagonal long-range order
in the ground state of a system of Bose particles confined by an external
potential in the dilute limit in which the Gross-Pitaewski equation becomes
exact. S\"uto shows condensation for the trapped gas at non zero temperature [4]. In [5] Lauers, Verbeure and Zagrebnov prove the
existence of a Bose-Einstein phase transition for the homogeneous gas under the
assumption that there is an energy gap at the bottom of the one-particle
spectrum. However, to our knowledge, there is still no proof of Bose-Einstein    
condensation in the interacting gas when there is no trap and no gap. In the 
present work we revisit this venerable many-body problem with a new point of 
view, the technique of quantum Mayer graphs. 

In order to provide orientation and motivation for our approach we recall some
facts pertaining to the mean-field
\footnote{Here we use "mean field" in the sense of van der Waals, 
that is collective effects of a long range interaction. The same word is also used in the context of the dilute limit
[3] leading to the Gross-Pitaewski regime. In fact, in the latter case, the situation is the opposite: dominant effects of the
interaction are due to rarefied local binary collisions.} 
Bose gas at an heuristic level. The
Hamiltonian of $N$ Bose particles in a volume $\Lambda$ interacting with
a constant repulsive potential of strength $a/|\Lambda|,\;a>0$, reads
\be
H_{{\rm mf},N}= H_{0,N}+\frac{a}{|\Lambda|}\frac{N(N-1)}{2}
\label{0.1}
\ee
where $H_{0,N}$ is the total kinetic energy. It leads to the free energy density
\be
f_{\rm mf}(\beta,\rho)= f_{0}(\beta,\rho)+\frac{a}{2}\rho^{2}
\label{0.1a}
\ee
where $\rho$ is the particle density, $\beta$ is related to the temperature $T$ by 
 $\beta= 1/k_{B}T$, $k_{B}$  denotes the Boltzmann constant, 
and $f_{0}(\beta,\rho)$ is the free energy of the non-interacting
Bose gas. Differenciating with respect to $\rho$ yields the relation
\be
\mu_{0}(\beta,\rho)=\mu_{{\rm mf}}(\beta,\rho)-a\rho
\label{0.1b}
\ee
between the chemical potential $\mu_{0}(\beta,\rho)$ of the free gas and the chemical 
potential $\mu_{{\rm mf}}(\beta,\rho)$
of the mean-field gas as functions of the density. Since the grand canonical densities
 $\rho_{0}$ of the free gas
and the density $\rho_{{\rm mf}}$ of the mean-field gas, considered
as functions of the respective chemical potentials, are the inverse functions of 
$\mu_{0}(\beta,\rho)$ and $\mu_{{\rm mf}}(\beta,\rho)$ at fixed
$\beta$, (\ref{0.1b}) is equivalent to the self-consistent equation
\be 
\rho_{{\rm mf}}(\beta,\mu)=\rho_{0}(\beta,\mu-a\rho_{{\rm mf}}(\beta,\mu))
\label{0.2}
\ee  
In (\ref{0.2})
\be
\rho_{0}(\beta,\mu)=\frac{1}{(2\pi \lambda^{2})^{3/2}}\sum_{q=1}^{\infty}
\frac{e^{\beta\mu q}}{q^{3/2}}
\label{0.3}
\ee
which is the well known formula for the grand canonical density of the free gas with
\be
\lambda=\hbar\sqrt
{\beta/m}
\label{0.3a}   
\ee
representing the thermal wavelength [6]. 

The series (\ref{0.3}) converges
for $\mu\leq 0$, so that the self-consistent equation (\ref{0.2}) is
meaningful whenever $\nu\equiv\mu-a\rho_{{\rm mf}}(\beta,\mu) \leq 0$, namely
for $\mu\leq \mu_{c}$ where the critical chemical potential $\mu_{c}$ is given by

\[ \mu_{c}-a\rho_{{\rm mf}}(\beta,\mu_{c}) =0. \]
Thus $\mu_{c}$ has the value
$\mu_{c}=a\rho_{0,c}(\beta)$ where $\rho_{0,c}(\beta)=\rho_{0}(\beta,\mu=0)$ is the 
critical density of the free gas.

At this point it is worth noting that the solution $\rho_{{\rm mf}}(\mu)$
of (\ref{0.2}) for $\mu\leq \mu_{c}$ can be extended to the range
$\mu>\mu_{c}$ by continuity. Indeed, we differentiate equation (\ref{0.2})
with respect to the chemical potential to obtain
\footnote{We keep the temperature fixed and omit from now on
$\beta$ in the notation.} 
\be
(\rho_{{\rm mf}})'(\mu)=\frac{(\rho_{0})'(\nu)}{1+a(\rho_{0})'(\nu)}
\label{0.4}
\ee
One sees from (\ref{0.3}) that $(\rho_{0})'(\nu)=\infty$ for $\nu>0$. Hence 
$(\rho_{{\rm mf}})'(\mu)=1/a, \;\nu>0$
and requiring the continuity of the density at $\mu=\mu_{c}$ gives
\be 
\rho_{{\rm mf}}(\mu)= \frac{\mu}{a}, \quad\;\mu>\mu_{c}
\label{0.6}
\ee
Equations (\ref{0.2}) and (\ref{0.6}) define the density for all values of
the chemical potential.  For $\mu>\mu_{c}$, there is a Bose
condensate of density $\rho_{\rm mf}(\mu)-\rho_{0,c}$.
These facts have been established with full mathematical rigor in
several works [7, 8, 9] and see [10] for a review. In particular they are obtained for the Kac 
interparticle potential 
\be
\label{0.5a}
V_{\gamma}(\br)=\gamma^{3}V(\gamma\br)
\ee 
in the scaling limit of $\gamma $ tending to zero. When $\gamma\to 0$, the potential 
$V_{\gamma}(\br)$  extends its range to infinity whereas its amplitude tends to zero 
in such a way, that the mean potential energy
\be
 \int d\br \gamma^{3}V(\gamma\br) \; = \; a   \label{aparameter}
\ee
stays constant. It is well known that in classical statistical mechanics this is the appropriate limit 
to rigorously recover the van der Waals mean field theory [11]. Also the methods of Mayer graphs
enable to calculate corrections to the mean field limit for small $\gamma$ [12].

In this paper we propose a similar approach to study an interacting Bose gas with
non-singular repulsive interactions. In section 2 we recall the "polymer" 
representation
of the Bose gas in thermal equilibrium. Combining the Feynman-Kac functional integral 
representation 
of the Gibbs weight together with the decomposition of  permutations into cycles, 
one finds the grand-canonical partition function in the classical-like form of 
a gas of interacting
polymers. Polymers are Brownian closed loops associated with a number of Bose 
particles belonging to a permutation cycle. Each loop has a self energy and there 
is a loop-loop pair potential.
In the space of polymers, all the techniques of classical statistical mechanics
are available, in particular the analysis of the partition function and of the density 
with the Mayer diagrammatic techniques. In this way we show in section 3 how 
the mean field equation (\ref{0.2}) can be recovered by summing the tree graphs.

In section 4 we treat an interacting gas with a general short range repulsive two-body potential and establish  
that its density obeys an exact equation of the form
\be
\rho(\mu)=F(\mu-a\rho(\mu))
\label{0.7}
\ee
The function $F(\mu)$ is defined as the sum of multiply connected Mayer graphs. 
We prove with the help of the
Penrose tree-graph inequality that the corresponding diagrammatic expansion of $F$ 
is convergent at low 
density, namely if the chemical potential is sufficently negative. Equation (\ref{0.7})
provides a generalization of the mean-field equation (\ref{0.2}) to the 
interacting gas:
indeed if one introduces a Kac potential $V_{\gamma}(\br)$, it is seen that 
$F_{\gamma}(\mu)$ reduces to the density $\rho_{0}(\mu)$ of the free gas as 
$\gamma\to 0$, thus recovering equation (\ref{0.2}).
The central goal then is to extend the validity of equation (\ref{0.7}) to higher 
densities (hopefully up to a critical density), that is, to approach the Bose 
transition point from the dilute phase, as in the mean field theory.

In section 5 we discuss the mathematical problems that arise at this point. 
The study of the
critical point will require the control of the asymptotic behaviour of the partition 
function of a single long repulsive polymer as well as the mutual interactions 
between different polymers. One should note
however that the  polymers occuring in the representation of the Bose gas differ from 
the standard self-repelling classical polymers because of the specifically 
quantum mechanical "equal time interaction" introduced by the Feynman-Kac formula. 
Hence, the results
of the theory of classical polymers cannot be used without further consideration.
As first investigation in this direction, we give lower and upper bounds 
on the partition function of a single polymer indicating that equation (\ref{0.7}) 
continues to hold for a range of chemical
potentials larger than that assuring the convergence of the Mayer series. 
Moreover we can calculate
the effects of mutually interacting polymers at lowest order in
the Kac parameter $\gamma$ by summing up the ring diagrams in a closed form.
Concluding remarks are presented in section 6. Proofs of some lemmas are relegated to appendices
1 and 2, and appendix 3 is devoted to the extension of
our methods to an inhomogeneous Bose gas confined by an external potential.

\section{The polymer representation of the Bose gas}

We consider bosons of mass $m$ in three dimensions with Hamiltonian
\be
H_{N} = -\frac{\hbar^{2}}{2m}\sum_{i=1}^{N}\Delta_{i} +\sum_{0\leq i<j\leq N}
V(\br_{i}-\br_{j})
\label{2.1}
\ee
enclosed in a box $\Lambda$, $\Delta$ being the Laplacian with Dirichlet 
conditions at the boundary of $\Lambda$. The pair potential $V(\br)$ is short 
range, repulsive and without singularities:
\be
V(\br)\geq 0, \quad\quad  \int{d\br}V(\br)\equiv a <\infty, \quad\quad 
\sup_{\br}V(\br)\equiv\bar{V}<\infty
\label{2.2}
\ee
We assume moreover that its Fourier transform $\tilde{V}(\bk)$ is positive.  

The polymer representation of the  grand-partition function at chemical potential $\mu$ and inverse temperature $\beta$
arises when the Gibbs statistical weight is expressed in terms of the Feynman-Kac path integral where quantum fluctuations are
represented by Brownian trajectories. The open trajectories associated with exchange contributions are reorganized
in larger closed loops (or polymers) containing several particles according to
the decomposition of permutations into cycles. 
The result is that the grand partition function of the Bose gas can  be written
in a classical-like form as (the so-called magic formula)
\be
\Xi_\Lambda=\sum_{n=0}^\infty\frac{1}{n!}\int\prod_{i=1}^nd\cl_iz(\cl_i)
\exp[-\beta U(\cl_1,\ldots,\cl_n)]
\label{2.100}
\ee
provided that suitable definitions of the phase space integration and 
of the interaction are given.
In one form or another, this representation has been known since a long time 
in various 
contexts starting with the work of Ginibre on the convergence of quantum virial 
expansions [13]. It is also used to implement numerical simulations of 
the Bose gas [14].
The present form (\ref{2.100}) has been derived and applied by Cornu [15] 
to Coulomb systems and we follow here the definitions given in Chap.V of [16]. 
A self contained derivation can also be found in [17].

An element $\cl$ of the phase space, called a loop or a polymer, 
\be
\cl=(\bR,\;\;q,\;\;\bX(s),\;\;0\leq\;s\;\leq q)
\label{2.200}
\ee
is specified by its position  $\bR$ in space, the number $q$ of particles 
belonging to it
and its shape $\bX(s)$. The $q$ particles are located at positions
\be
\br_{k}=\bR + \lambda\bX(k-1), \quad k=1,\ldots,q, \quad q+1\equiv 1
\label{2.3}
\ee
and
\be
\label{2.4}
\br_{k,k+1}(s)=\bR +\lambda\bX(k-1+s),\quad 0\leq s \leq 1
\ee
is an open path joining the $k$ particle at $\br_{k}$ to the $k+1$ particle at $\br_{k+1}$
where $\lambda$ is the thermal wave length (\ref{0.3a}). 
The loop can be 
viewed as an extended object at $\bR$ that has internal degrees of freedom 
$(q,\bX)$ with $q$ the number of particles belonging to a permutation cycle 
and $\bX$ the shape of the loop,
see Fig. 1. 

\vspace{3mm}

\begin{figure}
\begin{center}
\label{Fig. 1}
\includegraphics[width=9cm]{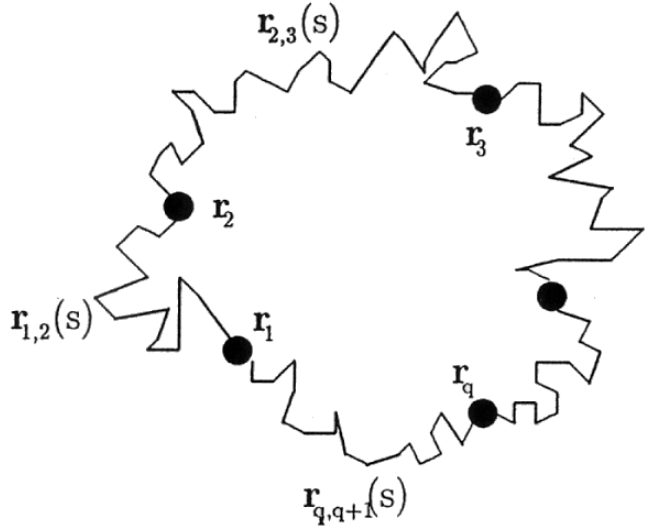}
\caption{A $q$ particle loop}
\end{center}
\end{figure}

\vspace{3mm}

The thermal wavelength $\lambda$ (the only place where the Planck constant 
occurs) gives the extent of quantum fluctuations. 
The shape of the loop $\bX$ is a Brownian bridge (a closed Brownian path), parametrized
by the "time" $s$ running in the "time interval" $[0,q]$ with $\bX(0)=\bX(q)=0$.
It is distributed according to the normalized Gaussian measure $\int D_{q}(\bX)\ldots$,  with covariance 
\be
\int D_{q}({\bf X }) X_{\mu} (s_{1})
X_{\nu}(s_{2})=\delta_{\mu\nu}q\left[\min\left(\frac{s_{1}}{q},\frac{s_{2}}{q}\right)
-\frac{s_{1}}{q}\frac{s_{2}}{q}\right]
\label{2.5}
\ee
where $X_{\mu}, \mu=1,2,3$ are the Cartesian coordinates of $\bX$.
Integration on phase space means integration over space and summation over all 
internal degrees of freedom of the loop
\be
\int d\cl\cdots= \sum_{q=1}^\infty
\int D_{q}(\bX)\int_{\Lambda} d\bR \cdots
\label{2.6}
\ee
Because of the Dirichlet boundary condition, the paths $\bX$ are constrained to
stay in the volume $\Lambda$,
but we do not write this constraint explicitly since it will be  
removed later in the infinite volume limit.

The interaction energy of two loops $\cl_i,\cl_j$ is the sum of pair Feynmam-Kac potentials between the particles
associated to the loops
\begin{eqnarray}
V(\cl_i,\cl_j) &
= & \sum_{k=1}^{q_{i}}\sum_{\ell =1}^{q_{j}}\int_{0}^{1} ds \,
V\left( \bR_i+\lambda\bX_{i}(k-1+s)-\bR_j-\lambda\bX_{j}(\ell -1+s)  \right)
\nonumber  \\
& = & \int_0^{q_{i}}ds_{i}\int_0^{q_{j}}ds_{j}{\tilde \delta}
(s_i-s_j)V(\bR_i+\lambda\bX_{i}(s_i)-\bR_j-\lambda\bX_{j}(s_j)) \nonumber \\
&& \label{2.7}
\end{eqnarray}
In (\ref{2.7}), the distribution
\be
{\tilde \delta}(s)=\sum_{n=-\infty}^{\infty}e^{2i\pi ns}
\label{2.8}
\ee
is the periodic Dirac function of period $1$.

The activity $z(\cl)$ of a loop is related to the chemical potential $\mu$ of the particles by 
\be
z(\cl)=
\frac{z^{q}}{q(2\pi q\lambda^2)^{3/2}} \;\exp(-\beta U(\cl)),\;\;\;z=e^{\beta \mu}
\label{2.11} 
\ee
It incorporates  the interactions $U(\cl)$ of the
particles in the same loop (the self energy of the loop)

\begin{eqnarray}
U(\cl)
=\frac{1}{2}\int_0^{q}
ds_1\int_0^{q}ds_2{\tilde \delta}(
s_1- s_2)V(\lambda(\bX(s_1)-\bX(s_2)))-\frac{1}{2}qV(0)\nonumber\\
\label{2.12}
\end{eqnarray}
The last term substracts out the self energy of the particles.
$U(\cl)$ can as well be written as
\be
U(\cl)
=\frac{1}{2}\int_0^{q}
ds_1\int_0^{q}ds_2{\tilde \delta}
(s_1- s_2)(1-\delta_{[s_{1}],[s_{2}]}^{Kr})
V(\lambda(\bX(s_1)-\bX(s_2)))\;\geq 0
\label{2.12b}
\ee
making manifest that $U(\cl)$ is a positive quantity (in the Kronecker delta symbol $[s]$ 
denotes the integer part of $s$). 
As a direct consequence one gets the bound
\be
0\leq z(\cl)\leq 
\frac{e^{\beta\mu q}}{q(2\pi q\lambda^2)^{3/2}}\equiv z^{(0)}(q)
\label{2.12a}
\ee

  From the structure (\ref{2.100}) of the partition function and the above definitions,
it is clear that the calculation rules of classical statistical mechanics apply to
the system of loops. We shall take advantage of this fact to analyze the system of loops
first and derive from there the results for the original quantum gas of particles.
In particular all the powerful techniques of Mayer graphs are available to
expand the loop density and the loop correlations in powers of the loop activities $z(\cl)$.
It is convenient to introduce the abbreviated notation $\cl_{i}=i$ and $d\cl_{i}=di$. 
Mayer bonds are defined by
\be
f(i,j)= e^{-\beta V(i,j)}-1
\label{2.13}
\ee
and weights at vertices by $z(i)$ (\ref{2.11}). Integration at vertices $di$ has to be 
performed according to (\ref{2.6}).
 
Notice that the bond is integrable over space since from the positivity of $V(\br)$, 
\be
|f(i,j)|\leq \beta V(i,j)
\label{2.14}
\ee
and from (\ref{2.7})
\beq
\int dR_{j}|f(\cl_{i},\cl_{j})|&\leq & \beta\int dR_{j}V(\cl_i,\cl_j)\nonumber\\
&=&\beta\int_0^{q_{i}}ds_{i}\int_0^{q_{j}}ds_{j}{\tilde \delta}(s_{i}-s_{j}) \int d\bR V(\bR) \nonumber \\
&=& \beta a q_{i}q_{j}
\label{2.15} 
\eeq
The loop density $\rho_{{\rm loop}}(\cl)$ is given by the standard expansion
\be
  \rho_{{\rm loop}}(1) = \sum_{n=1}^{\infty}nB_{n}(1)   \label{2.14a}
\ee
where
\be
B_{n}(1) = \sum_{n=1}^{\infty}\frac{1}{n!}\int d2\cdots dnz(1)z(2)\cdots
z(n)u(1,2,\ldots,n) \label{2.14aa}
\ee
and
\be
u(1,2,\ldots,n)=\sum_{\Gamma_{n}}\prod_{(i,j)\in \Gamma_{n}}f(i,j)
\label{2.15a}
\ee
is the Ursell function. The sum runs over all labelled connected graphs $\Gamma_{n}$ with $n$
vertices. We have directly written the loop density (\ref{2.14aa}) in the infinite volume 
limit, namely extending the spatial integrals $d\bR_{2}, \ldots d\bR_{n}$ at vertices 
$2, \ldots,n$  over the whole space,
whereas the vertex $1$ (the root point of the graph) carries no spatial integration. The 
existence of the
infinite volume limit of individual Mayer graphs follows from the translation invariance and 
integrability of the Mayer bond
as in the classical case. As a consequence $\rho_{{\rm loop}}(\cl)=\rho_{{\rm loop}}(q, \bX)$ 
does not depend
of the location $\bR$ of the loop $\cl$ in space.

Finally, to obtain the original particle density $\rho(\mu)$ from the loop density, we have to
 sum $\rho_{{\rm loop}}(\cl)$ over the internal variables of $\cl$,
\be
\rho(\mu)=\sum_{q=1}^{\infty}q\int D_{q}(\bX)\rho_{{\rm loop}}(q, \bX)
\label{2.16}
\ee
the aditionnal $q$ factor taking into account that the loop $\cl$ carries $q$ particles.

Although classical methods have been used, it is important to stress the difference between
the loop representation of the equilibrium state of the quantum Bose gas and that of
a gas of genuine classical polymers. 
First the chemical potential $\mu$ is not the variable thermodynamically conjugate to the
polymer number, but to the original particle number. Moreover, the loop interactions differ 
from the classical polymer
intractions by the quantum mechanical "equal time prescription" which originates from the 
Feynman-Kac formula.
This equal time prescription is manifested in (\ref{2.7}) and (\ref{2.12}) by the occurence 
of the periodic delta
function ${\tilde \delta}( s_1- s_2)$. In the classical interaction, 
every segment of a polymer interacts pairwise with any other segment, which would correspond 
to the potentials   
(\ref{2.7}) and (\ref{2.12}) without the equal time prescription.
The purely classical gas of point particles interacting by means of the two-body potential 
$V(\br)$ is recovered
if the thermal lenth $\lambda$ is set equal to zero and if only terms with $q=1$ (Boltzmann 
statistics) are retained.

\section{Mean field limit and tree graph summation}

Let us first show how the self-consistent mean field Bose gas is recovered in the 
diagrammatic analysis.
To this end we split the bond $f(i,j)$ into its  part linearized in the potential and higher 
order terms
\beq
f(i,j)&=&f^{(a)}(i,j) +f^{(b)}(i,j)\nonumber\\
f^{(a)}(i,j)&=&-\beta V(i,j) \nonumber\\ 
f^{(b)}(i,j)&=& e^{-\beta V(i,j)}-1 + \beta V(i,j)   
\label{3.1}
\eeq
These bonds are represented in Fig. 2.

\vspace{5mm}

\begin{figure}
\begin{center}
\label{Fig. 2}
\includegraphics[width=3cm]{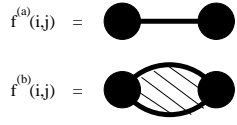}
\caption{The bonds $f^{(a)}(i,j)$ and $f^{(b)}(i,j)$ }
\end{center}
\end{figure}

\vspace{5mm}

\noindent 
This simply  enlarges the previous class of graphs $\Gamma$ to graphs (still 
denoted by $\Gamma$) where each bond can be
either $f^{(a)}(i,j)$ or $f^{(b)}(i,j)$.

In the mean field limit $\gamma\to 0$, the contributions of order ${\cal O}(1)$ in $\gamma$ 
due to the scaled potential $\gamma^{3}V(\gamma\br)$ will come from the linearized bonds
\be
f_{\gamma}^{(a)}(i,j)=-\beta \gamma^{3}
\int_0^{q_{i}}ds_{i}\int_0^{q_{j}}ds_{j}{\tilde \delta}(
s_i-s_j)V(\gamma(\bR_i-\bR_j)+\gamma\lambda(\bX_{i}(s_i)-\bX_{j}(s_j)))
\label{3.2}
\ee
and from the vertex weights
\be
z^{(0)}(i)=\frac{z^{q_{i}}}{q_{i}(2\pi q_{i}\lambda^2)^{3/2}}
\label{3.3}
\ee
The bond $f_{\gamma}^{(a)}(i,j)$ will give contributions of order ${\cal O}(1)$ since 
under scaling (\ref{aparameter}) its total spatial integral
\be
\int d\bR_{j}f^{(a)}_{\gamma}(\cl_{i,j})=-\beta a  q_{i}q_{j}
\label{3.4}
\ee
is independant of $\gamma$. In the activity (\ref{2.11}) we simply 
disregard the self-energy $U_{\gamma}(\cl)$ since the latter is ${\cal O}(\gamma^{3})$.

\vspace{3mm}

{\bf Proposition 1}

\noindent{\it The density $\rho_{{\rm tree}}(\mu)$ calculated as the the sum of all tree 
graphs with bonds (\ref{3.2})
and vertices (\ref{3.3}) verifies the mean field equation (\ref{0.2}).}  

\vspace{3mm}

Consider a rooted tree graph $T_{n+1}$ with vertices $(0,1,\ldots,n)$ for which
the root point is of degree $1$ ($0$ is the label of the root point)  
\footnote{The degree of a point is the number of lines incident at this point.}, Fig. 3. Hence the 
root point is linked to the rest of the
graph by a single bond, say $f^{(a)}(0,1)$.
Call $T_{n}$ the subgraph of $T_{n+1}$ with vertices $(1,2,\ldots, n)$ and $t_{n}(1)$ the 
value of this subgraph once integrated
on the vertices $2,\ldots,n$. Then the value $t_{n+1}(\cl_{0})$ of the rooted
graph $T_{n+1}$ is  
\be
t_{n+1}(\cl_{0})=-\beta z^{(0)}(q_{0})\int d\cl_{1}V_{\gamma}(\cl_{0},\cl_{1})t_{n}(\cl_{1})
\label{3.5}
\ee

\vspace{3mm}

\begin{figure}
\begin{center}
\label{Fig. 3}
\includegraphics[width=4cm]{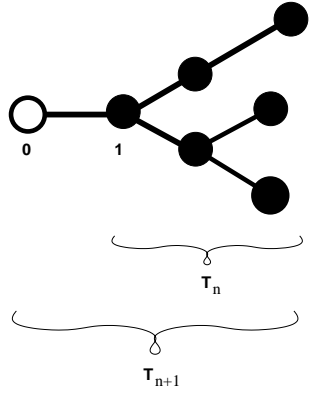}
\caption{Rooted tree $T_{n+1}$ with subtree $T_{n}$}
\end{center}
\end{figure}

\vspace{3mm}

\noindent The root point $0$ can be linked to any of the n vertices of the graph $T_{n}$. The resulting
tree graphs are different beacuse they contain different links with the root point.
Moreover, because of translational invariance $t_{n}(1)=t_{n}(q_{1},\bX_{1})$ does not depend
on the position $\bR_{1}$ of the loop $\cl_{1}$. 
Thus the spatial integration over $\bR_{1}$ can be performed on $V_{\gamma}(\cl_{0},\cl_{1})$ as 
in (\ref{3.4}) so that the total contribution reads
\be
nt_{n+1}(\cl_{0})=  z^{(0)}(q_{0})(-\beta aq_{0})\left[\sum_{q_{1}=1}^{\infty}q_{1}\int 
D(\bX_{1})nt_{n}(q_{1},\bX_{1})\right]
\label{3.6}
\ee
According to equations (\ref{2.14a}) and (\ref{2.16}) the quantity in the bracket 
is precisely the contribution to the particle density  
$\rho_{{\rm tree}}(\mu)$ of the graph $T_{n}$. Therefore the sum of all tree graphs having 
a root point of degree $1$
is $ z^{(0)}(q_{0})[-\beta aq_{0}\rho_{{\rm tree}}(\mu))$. The sum of all tree graphs with 
root point of degree $n$
is  $ z^{(0)}(q_{0})(-\beta a q_{0}\rho_{{\rm tree}}(\mu)]^{n}/n!$ (the factor $1/n!$ 
takes care of the fact that
the labelling of vertices belonging to different branches attached to the root point 
can be permuted without giving rise to new
Mayer graphs). Finally, summing on all trees rooted at the point $0$ gives the density of 
loops.  According to (\ref{2.16}), to obtain the particle density 
we still have to sum on the internal variables $q_{0}, \bX_{0}$ of the root loop 
with a factor $q_{0}$.
Hence using (\ref{3.3}) we find 
\beq
\rho_{{\rm tree}}(\mu)&=&
\sum_{q_{0}=1}^{\infty}\sum_{n=0}^{\infty}\frac{q_{0}z^{(0)}(q_{0})}{n!}
\(-\beta a q_{0}\rho_{{\rm tree}}(\mu)\)^{n}\nonumber\\
&=&\sum_{q_{0}=1}^{\infty}
\frac{\exp(\beta(\mu- a\rho_{{\rm tree}}(\mu)))}{(2\pi q_{0}\lambda^2)^{3/2}}\nonumber\\
&=&\rho_{0}(\mu- a\rho_{{\rm tree}}(\mu))
\label{3.7}
\eeq
which is the mean field equation (\ref{0.2}).

\vspace{3mm}

We now come back to the general Mayer series (\ref{2.14aa}) and prove its convergence 
at low density.

\vspace{3mm}

{\bf Proposition 2}

\noindent {\it The Mayer series (\ref{2.14aa}) converges for $\mu\leq -a\rho_{0,c}$ with 
$\rho_{0,c}=\rho_{0}(\beta ,\mu = 0)$ the critical density of the free gas.} 

\vspace{3mm}

We use the Penrose tree graph inequality for positive potentials which states that the  
the sum of the tree graphs provides an upper bound for the Ursell function [18]
\beq 
u(1,2,\ldots,n)&\leq &\sum_{T_{n}}\prod_{(i.j)\in T_{n}}|f(i,j)|\nonumber\\
&\leq &\sum_{T_{n}}\prod_{(i.j)\in T_{n}}(\beta V(i,j))
\label{3.8}
\eeq
the second inequality being a consequence of (\ref{2.14}). Moreover, if in the series 
(\ref{2.14aa}) we use the inequality (\ref{2.12a}),
we see that the tree summation with bonds 
$|f^{(a)}(i,j)|=\beta V(i,j)$ and vertices $z^{(0)}(i)$ provides also an upper bound.
According to the Proposition 1 the latter series sums up to a function 
$\bar{\rho}(\mu)$  that obeys 
\be
\bar{\rho}(\mu)=\rho_{0}(\mu + a\bar{\rho}(\mu))
\label{3.9}
\ee
This is the mean field equation of a Bose system for the negative potential $-V(\br)$: 
It has a finite solution provided $\mu + v\bar{\rho}(\mu)\leq 0$ which is equivalent to
$\mu\leq -\beta a \rho_{0,c}$.
We conclude that
\be
\rho (\mu) \leq\bar{\rho}(\mu))<\infty,\quad\quad \mu\leq - a \rho_{0,c}
\label{3.10}
\ee

  Notice that in the case of the scaled potential $V_{\gamma}(\br)$, the 
convergence of the series (\ref{2.14aa}) defining $\rho_{\gamma}(\mu)$ is not only absolute 
but also uniform with respect to $\gamma$ . This follows from the fact that in the evaluation 
of the tree-graph contributions one encounters only the integrated bonds 
        
\[ \int d\bR_{j}|f^{(a)}_{\gamma}(\cl_{i,j})|=\beta a  q_{i}q_{j} \] 

The Penrose inequality yields thus a $\gamma$-independent upper bound.

\vspace{3mm}

Now we show that in Kac limit, the density converges to the mean-field value.

\vspace{3mm}

{\bf Proposition 3}

\noindent {\it Let $\rho_{\gamma}(\mu)$ be the density associated with the scaled potential
$V_{\gamma}(\br)=\gamma^{3}V(\gamma\br)$. Then}
\be
\lim_{\gamma\to 0}\rho_{\gamma}(\mu)=\rho_{{\rm mf}}(\mu),\quad\quad \mu\leq -a\rho_{0,c}
\label{3.11}
\ee

\vspace{3mm}

Consider first a graph $\Gamma_{n}$ with vertices $1,2,\ldots,n$ (rooted at point $1$) which 
is not a tree, i.e 
it contains at least one cycle. Delete some bonds in such a way that the graph 
thus obtained is a connected tree $T_{n}$ with value $t_{n}(1)$.
In order to find an upper bound for the value $g_{n,\gamma}(1)$ of $\Gamma_{n}$ 
we use the inequalites 
\be
|f_{\gamma}(i,j)|\leq \beta V_{\gamma}(i,j)
\label{3.12}
\ee
for the bonds remaining in the tree and
\be
|f_{\gamma}(i,j)|\leq\gamma^{3}\beta q_{i}q_{j}\bar{V}
\label{3.13}
\ee
for each deleted bond.
Notice that (\ref{3.13}) is obtained by replacing the pair potential  in 
(\ref{3.2}) by its supremum $\bar{V}$. Moreover the activity $z(i)$ is bounded
by $z^{(0)}(i)$ (\ref{2.12a}). Spatial integration of a bond $(i,j)$ of the tree yields 
the factor $\beta a q_{i}q_{j}$ as before so
that with (\ref{3.13}) a vertex $(i)$ of degree $k_{i}$ in $\Gamma_{n}$  receives a 
factor $q_{i}^{k_{i}}$ multiplied by the activity $z^{(0)}(q_{i})$. If $\ell$ bonds have
been deleted, the resulting inequality has the form
\be
|g_{n,\gamma}(1)|\leq (\gamma^{3}\bar{V})^{\ell}(\beta a)^{n-1}\sum_{q_{2},\ldots, q_{n}=1}^{\infty}
q_{1}^{k_{1}}\cdots q_{n}^{k_{n}}z^{(0)}(q_{1})\cdots z^{(0)}(q_{n})
\label{3.15}
\ee
In view of the exponential factor $e^{\beta \mu q}$ in (\ref{2.12a}), the $q$-series in 
(\ref{3.15}) is convergent for $\mu < 0$.
Hence the presence of one cycle in $\Gamma_{n}$ implies that $g_{n,\gamma}(1)$ tends to zero 
not slower than $\gamma^{3}$ for $\gamma\to 0$ (and as $\gamma^{3\ell}$ if $\Gamma_{n}$ 
has $\ell$ cycles).

If $\Gamma_{n}$ is a tree we decompose the Mayer bonds as in (\ref{3.1}). Since
\be
\label{3.16}
|f^{(b)}(i,j)|\leq \frac{1}{2}\(V(i,j)\)^{2}\leq\frac{1}{2}\gamma^{3}\beta
q_{i}q_{j}\bar{V}(\beta V(i,j))
\ee
all the contributions to $\Gamma_{n}$ containing $f^{(b)}(i,j)$ bonds vanish as 
$\gamma\to 0$.   

As we have seen from the Penrose estimate and from Proposition 2, the Mayer series 
constituting $\rho_{\gamma}(\mu)$ 
is absolutly convergent uniformly with respect to $\gamma$. By dominated convergence, 
the limit of the diagrammatic sum can be calculated term by term. Therefore
we are left with the sum of trees where all the bonds are of the type $f^{(a)}_{\gamma}(i,j)$ 
and where the activites $z_{\gamma}(i)$
can be replaced by $z^{(0)}(i)$ as $\gamma\to 0$. This is exactly the situation of the 
Proposition 1 thus proving Proposition 3.

\section { The self-consistent equation for the interacting gas}

The reasoning which led to equation (\ref{3.7}) was restricted to the tree diagrams.
We now show how it can be generalized to provide an implicit equation for the exact 
density $\rho (\mu )$. 
To this end we consider the complete set of expanded Mayer graphs with the two types of bonds 
$f^{(a)}(i,j)$ and $f^{(b)}(i,j)$ defined in (\ref{3.1}) and
call $f^{(a)}(i,j)$ a single interaction bond. A graph is said multiply connected 
\footnote{In the context of Feynman diagrams, such graphs are also called one-line irreducibles.} 
if it cannot be disconnected by cutting a single interaction  bond. 
Then we define $I(\cl)$ to be the value of the sum of all
multiply connected graphs with  one root point $\cl$, see Fig. 4.

\vspace{3mm}

\begin{figure}
\begin{center}
\label{Fig. 4}
\includegraphics[width=13cm]{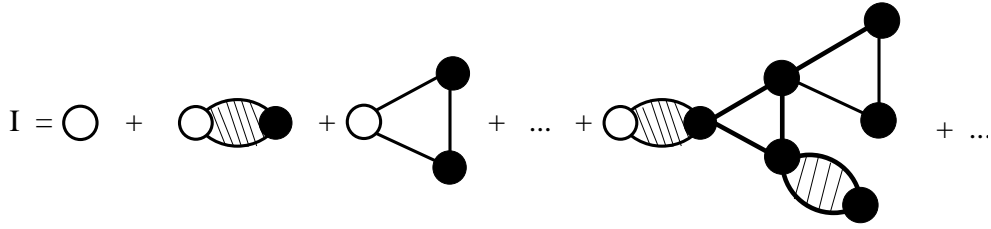}
\caption{The sum of multiply connected graphs}
\end{center}
\end{figure}

\vspace{3mm}

Because of translation invariance, $I(\cl)$ does not depend on 
the position $\bR$ of $\cl$. Denoting here the chemical potential by $\nu$~, we define 
the function
\be
\label{4.1}
F(\nu)=\sum_{q=1}^{\infty}q\int D_{q}(\bX)I(\cl)
\ee
For $\nu\leq -a\rho_{0,c}$, the function $F(\nu)$ is represented by a convergent sum
since it is a subseries of the absolutly convergent Mayer expansion (Proposition 2).

\vspace{3mm}

{\bf Proposition 4}
\noindent{\it For $\mu\leq - a \rho_{0,c}$, $\rho(\mu)$ verifies the equation}
\be
\label{4.2}
\rho(\mu)=F(\mu- a\rho(\mu))
\ee

To establish (\ref{4.1}) and (\ref{4.2}), we proceed as in Proposition 1. 
Considering now a general structure of Mayer graphs we say that a vertex $(i)$ in a graph 
has a star structure if there is a number of incident single interaction lines
at $(i)$ such that cutting any one of them disconnects the graph.
Consider first a vertex $(i)$ having a star structure consisting of one single interaction 
line. This interaction line, say $f^{(a)}(i,1)$, links the vertex $(i)$ to a subgraph 
$\Gamma_{n}$ with vertices
$(1,2,\ldots, n)$; $\Gamma_{n}$ has no other links with the rest of the graph, Fig. 5.  
Once integrated on the points $(2,\ldots,n)$, its value $g_{n}(1)$ does not depend on 
the position $\bR_{1}$ of the loop $\cl_{1}$. Thus the spatial integration over $\bR_{1}$ 
can be performed on the bond $[ -\beta V(\cl_{i},\cl_{1})]$ yielding the factor 
$(-\beta a q_{i}q_{1})$.
Taking finally into account that the graph $\Gamma_{n}$
can be attached through a single interaction line to the vertex $(i)$ 
by any of its $n$ vertices we find (as in (\ref{3.5})) that
its total  contribution to the vertex $(i)$ equals
\be
z(\cl_{i})(-\beta aq_{i})\left[\sum_{q_{1}=1}^{\infty}q_{1}\int
D_{q_{1}}(\bX_{1})\, ng_{n}(q_{1},\bX_{1})\right]
\label{4.3}
\ee
But from (\ref{2.16}), the quantity in the square bracket is the contribution of the graph 
$\Gamma_{n}$ to the exact density. 
Thus the sum of all such graphs contributes to the vertex $(i)$ as   
$z(\cl_{i})(-\beta a q_{i}\rho(\mu))$.  Continuing the reasoning along the lines 
of the proof of Proposition 1 we consider next the sum all star structures at $(i)$ 
having $n$ single interaction lines. Their contribution equals
$z(\cl_{i})[-\beta a q_{i}\rho(\mu)]^{n}/n! $. Finally, summing all possible star 
structures at $(i)$ we arrive at the formula for an effective activity
\be
z^{*}(\cl_{i})=z(\cl_{i})e^{-\beta a \rho(\mu) q_{i}}=
\frac{e^{\beta(\mu-a\rho(\mu))q_{i}}}{q_{i}(2\pi q_{i}\lambda^{2})^{3/2}}
e^{-\beta U(\cl_{i})}
\label{4.4}
\ee
This is precisely the previous activity $z(\cl_{i})$ (\ref{2.11}) evaluated at the shifted 
chemical potential $\nu= \mu -a \rho(\mu)$.

\vspace{3mm}

\begin{figure}
\begin{center}
\label{Fig. 5}
\includegraphics[width=5cm]{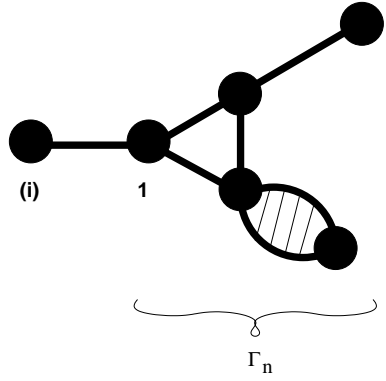}
\caption{Star structure at vertex (i) and subgraph $\Gamma_{n}$}
\end{center}
\end{figure}

\vspace{3mm}

>From this analysis we see that $\rho(\mu)$ can as well be  obtained 
by summing all multiply connected graphs with bonds $f^{(a)}(i,j)$ and $f^{(b)}(i,j)$ with 
effective activities $z^{*}(\cl_{i})$ at vertices.
 Indeed, by construction, any graph in these series is an original Mayer graph.
Conversely, each Mayer graph does appear therein. This can be shown by the following 
reasoning. Consider any connected graph made of 
the bonds $f^{(a)}(i,j)$ and $f^{(b)}(i,j)$ and remove all single interaction lines
$f^{a}(i,j)$ whose presence makes it not mutliply connected (by definition removal of such a 
line disconnects the root point from a part of the graph). Then the remaining subgraph 
connected to the root point is a multiply connected graph from which the original Mayer 
graph is obtained in a unique way by forming star structures. 
This concludes the proof since, in view of formula (\ref{4.4}), by forming $F(\nu)$ 
according to these rules we find the density equal to the value of this function at the 
shifted argument $\nu=\mu -a \rho(\mu)$. 

As a corollary of Propositon 3, we have that if $F_{\gamma}(\nu)$ is calculated with 
the scaled potential $V_{\gamma}(\br)$ then $\lim_{\gamma\to 0}F_{\gamma}(\nu)=
\rho_{0}(\nu)$.
   
 We can write $I(\cl)$ appearing in the definition (\ref{4.1}) of function $F_{\gamma}(\nu)$ 
as the sum  
\be
I(\cl)=z(\cl)[1+I_{\rm cycle}(\cl)]
\label{4.5}
\ee
of the pure root point term plus all the multiply connected graphs containing cycles and  
$f^{(b)}(i,j)$ bonds. It is clear from the proof of
Proposition 3 that the mean field density arises solely from the root point 
$z(\cl)$ whereas $I_{\rm cycle}(\cl)$ becomes vanishingly small in the mean field limit.

All the results presented so far are valid within the convergence radius of the Mayer 
series determined by $\mu<-a\rho_{0,c}$. The
question then arises if the density, as solution of (\ref{4.2}),
can be extended to larger values of $\mu$, as it was possible in the strict mean field case. 
An investigation of this question and of 
the mathematical difficulties involved is presented in the next section.

\section{ The Bose gas beyond mean field}

{\bf \large 5.1  Vertex contribution: the single polymer partition function}

\vspace{3mm}

In this section we study the structure of function $F(\nu)$ in a more detailed way.
The first term $f^{(0)}(\nu)$  of the expansion of $F(\nu)$ in multiply connected graphs 
corresponds to the root point $z(\cl)$ which is a single vertex contribution
\beq
f^{(0)}(\nu)&=&\sum_{q=1}^{\infty}q\int D_q(\bX)z(\cl)\nonumber\\
&=&\frac{1}{(2\pi \lambda^{2})^{3/2}}\sum_{q=1}^{\infty}\frac{e^{\beta\nu q}}{q^{3/2}}\kappa(q)
\label{5.1}
\eeq
Here we have introduced the mean value of the Boltzmann factor  
\be
\kappa(q)= < e^{-\beta U} >_{q} \, \equiv \, \int D_{q}(\bX)e^{-\beta U(q, \bX)}
\label{5.2}
\ee
of a single closed polymer carrying the repulsive energy (\ref{2.12}). The Brownian bridge measure defining 
the average $< ... >_{q}$ is normalized to $1$, namely if $U(q, \bX)=0$, one has $\kappa(q)=1$ 
and $f^{(0)}(\nu)$ reduces then to the free gas density (\ref{0.3}).
Moreover, for positive potentials we have $0\leq\kappa(q)\leq 1$. Clearly, 
the radius of convergence of the series ( \ref{5.1}) is determined by the asymptotic  
behaviour of $\kappa(q)$ for $q\to\infty$. The following bounds can be established

\vspace{3mm}

{\bf Lemma 1}

\noindent{\it There exists $\nu_{+}$ such that}
\be 
\kappa(q)\geq e^{-\beta q\nu_{+}}
\label{5.3}
\ee

This lower bound follows from Jensen inequality
\be
\kappa(q)\geq\exp(-\beta\langle U\rangle_{q})
\label{5.4}
\ee
where
\be
\langle U\rangle_{q}=\int D_{q}(\bX)U(q,\bX)
\label{5.5}
\ee
is the average energy of the polymer. The value of $\nu_{+}$ results from a direct 
calculation of $\langle U\rangle_{q}$ 
presented in Appendix 1.

\vspace{3mm}

There is also an upper bound:

\vspace{3mm}

{\bf Lemma 2}

\noindent {\it Let $r$ be a fixed integer $r\geq 2$. There exists $\nu_{-}$ (depending on $r$) such that $0<\nu_{-}<\nu_{+}$
and}
\be
\kappa(qr)\leq r^{3/2}\exp(-\beta qr\nu_{-})
\label{5.6}
\ee

The upper bound is derived by splitting the closed loop 
$\bX(s),\,0\leq s\leq q,\;\bX(0)=\bX(q)=0$, into the union
of two open Brownian paths  $\bX_{1}(s),\,0\leq s\leq q_{1}$ and 
$\bX_{2}(s),\,q_{1}\leq s\leq q$, and disregarding 
the (positive) interactions between the two paths $\bX_{1}$ and $\bX_{2}$. 
The details and the form of $\nu_{-}$ are given in Appendix 2.

The lemmas imply that $f^{(0)}(\nu)$ is finite if $\nu\leq\nu_{-}$ and diverges for $\nu>\nu_{+}$. 
Let us suppose for a moment that there exists a critical value $\nu_{c}$ such that
\beq
f^{(0)}(\nu)&<&\infty,\quad \nu\leq\nu_{c}\nonumber\\
f^{(0)}(\nu)&=&\infty, \quad  \nu > \nu_{c}
\label{5.7}
\eeq
In this case, there will be a critical density $\rho_{c}=f^{(0)}(\nu_{c})$ and a critical 
chemical potential $\mu_{c}= a\rho_{c}+\nu_{c}$. From the lemmas 1 and 2, one has necessarily
$\nu_{-}\leq \nu_{c}<\nu_{+}$. 

The determination of the possible existence of $\nu_c$ requires the knowledge of
the exact asymptotic behaviour of $\kappa(q)$ as $q\to\infty$. In this respect
let us make the following comment concerning the theory of classical polymers, defined as 
those interacting via the standard pairwise repulsion
\begin{eqnarray}
U_{\rm cl}(\cl)=
\frac{1}{2}\int_0^{q}
ds_1\int_0^{q}ds_2
V(\bX(s_1)-\bX(s_2))
\label{5.8}
\end{eqnarray}
It is firmly established, although not rigorously proved, that the normalized partition
function of a single classical closed polymer behaves (in three dimensions) as [19]
\be
\kappa_{\rm cl}(q)=\int D_{q}(\bX) e^{-\beta U_{\rm cl}(\cl) }\sim C\frac{e^{-\beta Aq}}{q^{3(\nu_{\rm pol}-1/2)}}\;,
\quad q\to\infty
\label{5.9}
\ee
where $A$ is a constant depending of the choice of the potential $V(\br)$ and $\nu
_{\rm pol}=0,589$ is the universal
critical exponent for a swollen polymer. Inserting this asymptotic behaviour in the series (\ref{5.1})
gives $\nu_{c}=A$ and a finite critical density $\rho_{c}$ since the series is convergent at 
$\nu~=\nu_{c}$. It is an open question to find out whether a similar situation holds for 
the "quantum" polymers  subjected to the "equal time" interaction (\ref{2.12}).

\vspace{3mm}

{\bf \large 5.2  Bond contributions: interacting polymers}

\vspace{3mm}

Interaction between different polymers occurs in multiply connected graphs having bonds 
$f^{(a)}(i,j)$ or $f^{(b)}(i,j)$.
It turns out that the subseries $I_{\rm ring}(\cl)$ of $I_{\rm cycle}(\cl)$ defined 
as the sum of all multiply connected
graphs having exactly one cycle of interaction bonds can be summed in a closed form,
see Fig. 6 (the first term of the series corresponds to the
quadratic term $V^{2}(i,j)/2$ in the expansion of $f^{(b)}(i,j)$,  (\ref{3.1})).
The result is
\be
I_{\rm ring}(\cl)=\frac{1}{2}\int d\bk \int_{0}^{q}ds \int_{0}^{q}dt 
e^{i\lambda\bk\cdot(\bX(s)-\bX(t))}
\sum_{n=-\infty}^{\infty}\frac{(\beta\tilde{V}(\bk))^{2}\alpha_{n}^{2}(\bk)}{1+\beta\tilde{V}
(\bk)\alpha_{n}^{2}(\bk)}
e^{2i\pi n(s-t)}
\label{5.10}
\ee
Here $\tilde{V}(\bk)$ is the Fourier transform of the potential and the positive coefficients 
 $\alpha_{n}^{2}(\bk)$ 
\be
\alpha_{n}^{2}(\bk)=\sum_{q=0}^{\infty}q\int_{0}^{q}ds\int D_{q}(\bX)z(\cl)
e^{i\lambda\bk\cdot\bX(s)}e^{2i\pi ns}
\label{5.11}
\ee
come from the summation on the internal degrees of freedom of loops at vertices.
A derivation of formula (\ref{5.10}) can be found in [20] 
where the effective
loop-loop potential is calculated as the sum of all chain graphs. 
$I_{\rm ring}(\cl)$ is given precisely by the formula (84) in
[20]  restricted to a single species of bosonic particles and with the
Fourier transform $4\pi/|\bk|^{2}$ of the Coulomb potential replaced by
the present short range potential $\tilde{V}(\bk)$ (in [20] the coefficients 
$4\pi \beta\alpha_{n}^{2}(\bk)$ are noted $\kappa^{2}(\bk, n)$).

\vspace{3mm}

\begin{figure}
\begin{center}
\label{Fig. 6}
\includegraphics[width=12cm]{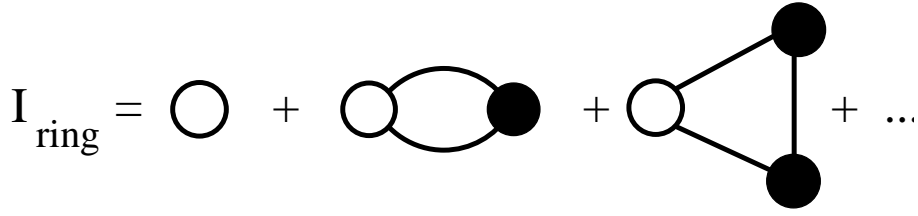}
\caption{The sum of rings}
\end{center}
\end{figure}

\vspace{3mm}

We now check that $I_{\rm ring}(\cl)$ is also well defined in the range $\nu<\nu_-$ 
(or possibly $\nu<\nu_c$ if there is a
critical value $\nu_c$ (\ref{5.7})). As the modulus of the phase factors in (\ref{5.10}) 
equals $1$ the positivity of
$\beta\tilde{V}(\bk)\alpha_{n}^{2}(\bk)$ implies the inequality
\beq
I_{\rm ring}(\cl)\leq\frac{q^{2}}{2}\int d\bk(\beta \tilde{V}(\bk))
\sum_{n=-\infty}^{\infty}\alpha_{n}^{2}(\bk)
\label{5.11a}
\eeq
>From (\ref{5.11}) and (\ref{2.8})
\beq
\sum_{n=-\infty}^{\infty}\alpha_{n}^{2}(\bk)=\sum_{q=1}^{\infty}q\int D_{q}(\bX)z(\cl)
\left[\int_{0}^{q}ds e^{i\lambda\bk\cdot\bX(s)}\tilde{\delta}(s)\right]
\leq f^{(1)}(\nu)
\label{5.12}
\eeq
We have defined
\beq
f^{(k)}(\nu)=\sum_{q=1}^{\infty}q^{k+1}\int D_{q}(\bX)z(\cl)
=\sum_{q=1}^{\infty}q^{k}\frac{e^{\beta\nu q}}{(2\pi q\lambda^{2})^{3/2}}\kappa(q)
\label{5.13}
\eeq
and the inequality follows from the fact that the bracket $[\cdots]$ is less or equal to $q$~.
Hence the contribution of the ring diagrams $ F_{\rm ring}(\nu)$ to the function $ F(\nu)$,
\be
F_{\rm ring}(\nu)=\sum_{q=1}^{\infty}q\int D(\bX)z(\cl)I_{\rm ring}(\cl)
\label{5.14}
\ee
is bounded by
\be
F_{\rm ring}(\nu)\leq\frac{1}{2}\int d\bk(\beta \tilde{V}(\bk))^{2}f^{(1)}(\nu)f^{(2)}(\nu)
\label{5.16}
\ee
where the inequality (\ref{2.12a}) has been also used.
In view of the Lemma $1$, it is clear that $ f^{(k)}(\nu)$ (\ref{5.13}) is finite for 
$\nu<\nu_-$
and thus $F_{\rm ring}(\nu)$ is also well defined in this extended range of
chemical potentials.

The ring contribution is expected to be small for a scaled potential $V_{\gamma}$ when 
$\gamma\to 0$. Indeed if the Fourier transform
$\tilde{V}_{\gamma}(\bk)=\tilde{V}(\bk/\gamma)$ of $V_{\gamma}$ is introduced in 
(\ref{5.16}) and the
$\bk$ integration variable is changed into $\bk=\gamma\bp$ we find immediately
\be
F_{\rm ring,\gamma}(\nu)\leq\frac{\gamma^{3}}{2}\int d\bp(\beta \tilde{V}(\bp))^{2}
f_{\gamma}^{(1)}(\nu)f_{\gamma}^{(2)}(\nu)
={\cal O}(\gamma^{3}),\quad \nu<\nu_-
\label{5.17}
\ee
In (\ref{5.17}), $f_{\gamma}^{(k)}(\nu)$ are still dependent on $\gamma$ through the 
single polymer partition function
$\kappa_{\gamma}(q)$.

If one introduces the scaled potential in the full expression (\ref{5.10}) 
(setting $\bk=\gamma\bp$) one observes
the subtle dependence of the parameter $\gamma$. 
There is an overall $\gamma^{3}$ prefactor that manifests the smallness of the scaled 
potential as in the classical case.
A $\gamma$ factor occurs also in the phase in the integrands of (\ref{5.10}) and 
(\ref{5.11}) in the combination
$\gamma\lambda$, representing the ratio of the thermal wave length and the potential range.
The importance of these phases will therefore  depend of the value of this ratio. 
Finally, $\gamma$ appears in the
loop activity $z_{\gamma}(\cl)$ in (\ref{5.11}).
For $\gamma$ small and $\gamma\lambda\ll 1$ we can replace the above mentionned phase 
factors by $1$ and approximate
\be
F_{\rm ring}(\nu)\sim \frac{\gamma^{3}}{2}f^{(2)}(\nu)
\int d\bp\frac{(\beta \tilde{V}(\bp))^{2}f_{\gamma}^{(1)}(\nu)}
{1+\beta \tilde{V}(\bp)f_{\gamma}^{(1)}(\nu)},\quad \gamma\to 0
\label{5.18}
\ee
This will be the dominant term in the expansion of $I_{\rm cycle}(\nu)$ since terms 
with $k>1$ cycles will obtain
an overall $\gamma^{3k}$  prefactor.

\section{Concluding remarks}

The use of quantum Mayer graphs proved useful in discovering the existence of an
implicit equation (\ref{4.2}) defining the density of an interacting Bose gas as 
function of its chemical potential. The knowledge of the precise form of the equation
requires the resummation of all multiply connected diagrams including the single vertex 
contribution. 
It should be recalled that in a series of papers, Lee and Yang have developped a diagrammatic formalism for the quantum statistical
mechanical many-body problem that enables to calculate the thermodynamical quantities in terms of Boltzmann-type
Ursell functions together with rules taking quantum statistics into account (see [21] for the general formalism and [22]
for application to bosons). Working in the occupation number representation in momentum space, these authors also obtain a formally
exact integral equation for the average occupation of modes, and in particular for the condensate density.
The latter equation (a generalization of the Bogoliubov condensate equation) was shown to be exact in the thermodynamic limit by
Ginibre [23], and it has also been established in the framework of infinitely extended states of Bose systems [24].
Our formulation is different in the sense that it provides a closed equation for the density of the interacting gas
at fixed chemical potential that is particularly well adapted to the exploration of the neighborhood
of the mean field limit. Moreover it provides an interesting link with the theory of polymers.

Suppose that the resummation of multiply connected diagrams leading to (\ref{4.2}) can been performed. It is then not excluded that 
the resulting non-perturbative equation remains valid beyond the radius of 
convergence of the Mayer series. If so, the question of the Bose-Einstein condensation in 
an interacting gas could be examined on the basis of (\ref{4.2}).
 At least one example supports this hope: such an extension to the transition region 
agrees with rigorous results in the case of the  mean field limit.
 This fact motivates the study of the self-consistent relation (\ref{4.2}) beyond the mean 
field limit.  We formulated this problem here using the scaling (\ref{0.5a}) of the pair 
potential. In order to derive the form of function $F_{\gamma}$ defining the relation 
(\ref{4.2}) for the scaled potential one needs to know  the small $\gamma$ asymptotics of
\begin{itemize}
\item{(i)} the mean value of the Boltzmann factor of a single polymer 
(or normalized partition function)
\item{(ii)} the sum of ring diagrams representing the contribution of interacting polymers
\end{itemize}   
In the study of point (i) we could not use directly the known results (\ref{5.9})
because of equal time condition imposed by quantum mechanics on 
interactions between different elements of polymers (see (\ref{2.7})).
The existence and localization of the Bose-Einstein condensation depends in a crucial way 
on the behavior of (i) for extended polymers.
This however remains an open problem. Our paper provides only exact upper and lower 
bounds which, what is interesting, turn out to be qualitatively compatible with the classical  
result (\ref{5.9}). There are some indications that in the $\gamma\to 0$ limit the 
quantum calculation approaches the classical one. However, this question is at present 
not yet understood. 
We based the analysis of point (ii) on the remarkable fact that quantum Mayer ring diagrams
can be summed in closed analytical form. 
We thus arrived at an analytic expression (\ref{5.10}) for the sum of 
the relevant ring diagrams. However here again the small $\gamma$ limit of (\ref{5.10}) 
involves the normalized single polymer partition function whose asymptotics for extended 
polymers remains to be derived. We hope that the continuation of the study of 
quantum Mayer graphs along these lines and progress in the understanding of polymer partition functions will eventually lead to the
determination of  functional relation (\ref{4.2}) beyond the mean field theory.

\vspace{5mm}

{\Large Acknowledgments}

\vspace{3mm}

\noindent We thank B. Duplantier for a stimulating discussion on links with the theory of polymers.
One of us (JP) greatly acknowledge the hospitality at the \'Ecole Polytechnique F\'ed\'erale
de Lausanne where this research has been performed.

\vspace{5mm}

{\Large {\bf Appendices}}

\vspace{3mm}

\noindent {\bf \large Appendix 1 : Proof of lemma 1}

\vspace{3mm}

The average energy (\ref{5.5}) is calculated from its definition (\ref{2.12}) (introducing also the Fourier transform
$\tilde{V}(\bk)$ of the potential)
\be
\label{A.1}
\langle U\rangle_{q}=J(q)-q\frac{V(0)}{2}
\ee
where
\beq
J(q)&=&\frac{1}{2}\int D_{q}(\bX)\int_0^{q}
ds_1\int_0^{q}ds_2{\tilde \delta}(
s_1- s_2)V(\lambda(\bX(s_1)-\bX(s_2)))\nonumber\\
&=&\frac{1}{2}\int d\bk\tilde{V}(\bk)\int_0^{q} ds_1\int_0^{q}ds_2{\tilde \delta}(s_1- s_2)\int D_{q}(\bX)
e^{i\lambda\bk\cdot\left[\bX(s_{1}-\bX(s_{2})\right]}\nonumber\\
\label{A.2}
\eeq
>From the basic rules for Fourier transforms of Gaussian measures we have using the covariance (\ref{2.5})
\be
\int D_{q}(\bX)
e^{i\lambda\bk\cdot\left[\bX(s_{1}-\bX(s_{2})\right]}= \exp\left[-\frac{\lambda^{2}k^{2}}{2}C_{q}(s_{1}-s_{2})  \right],
\quad k=|\bk|
\label{A.3}
\ee
where
\be
C_{q}(s)=|s|\(1-\frac{|s|}{q}\right)
\label{A.4}
\ee
Since both $\tilde{\delta}(s)$ and $C_{q}(s)$ can be considered as periodic functions of period $q$ 
the double time integral in (\ref{A.2}) reduces to
\beq
& &\int_0^{q} ds_1\int_0^{q}ds_2{\tilde \delta}(s_1- s_2)\exp\left[-\frac{\lambda^{2}k^{2}}{2}C_{q}(s_{1}-s_{2})\right]\nonumber\\ 
&=&
q\sum_{n=0}^{q-1}\exp\left[-\frac{\lambda^{2}k^{2}}{2}C_{q}(n)\right]=q\sum_{n=1}^{q-1}
\exp\left[-\frac{\lambda^{2}k^{2}}{2}C_{q}(n)\right] 
-q \label{A.5}
\eeq
When this is introduced in (\ref{A.2}) and (\ref{A.1}) (noting that $\int d\bk\tilde{V}(\bk)=V(0)$) one obtains finally
\be
\langle U\rangle_{q}=q\int d\bk  \tilde{V}(\bk)g(k,q)
\label{A.6}
\ee
with
\be
g(k,q)=\frac{1}{2}\sum_{n=1}^{q-1}\exp\left[-\frac{\lambda^{2}k^{2}}{2}C_{q}(n)\right]
=\sum_{n=1}^{q/2}\exp\left[-\frac{\lambda^{2}k^{2}}{2}C_{q}(n)\right]
\label{A.7}
\ee
The last equality results from the symmetry $C_{q}(n)=C_{q}(q-n)$. If $n\leq q/2$, one has 
obviously $C_{q}(n)\geq n/2$ which leads to the bound
\be
g(k,q)\leq \sum_{n=1}^{q/2}\exp\left[-\frac{\lambda^{2}k^{2}}{4}n\right]\leq \frac{1}{\exp(\frac{\lambda^{2}k^{2}}{4})-1}
\label{A.8}
\ee
hence
\be
\langle U\rangle_{q}\leq q\nu_+,\quad\quad \nu_+ = \int d\bk \frac{\tilde{V}(\bk)}{\exp(\frac{\lambda^{2}k^{2}}{4})-1} 
\label{A.9}
\ee
thus proving Lemma 1.

It is instructive to exhibit the behaviour of $\nu_+$ in the scaling limit of a Kac potential
$\tilde{V}_{\gamma}(\bk)=\tilde{V}(\bk/\gamma)$. Setting $\bk =\gamma\bp$ in (\ref{A.9}) gives
\be
\nu_+=\gamma^{3}\int d\bp\frac{\tilde{V}(\bp)}{\exp(\frac{\gamma^{2}\lambda^{2}p^{2}}{4})-1} 
\sim \frac{4\gamma}{\lambda^{2}}\int d\bp\frac{\tilde{V}(\bp)}{p^{2}},\quad \gamma\to 0 
\label{A.10a}
\ee

\vspace{3mm}

\noindent {\bf \large Appendix 2 : Proof of lemma 2}

\vspace{3mm}

We denote
$\langle F\rangle_{W}=E(F|\bX(0)=0)$ the normalized Wiener expectation of a functional $F(\bX)$ of paths $\bX(t)$
starting from the origin at time $t=0$ and $E(F|\bX(t_{1})=\bR_{1},\bX(t_{2})=\bR_{2})$ the conditional Wiener expectation for paths
starting in $\bR_{1}$ at time $t_{1}$ and ending in $\bR_{2}$ at time $t_{2}$. One has in particular
\beq
&&\langle F\rangle_{W}=\int d\bR E(F|\bX(0)=0,\bX(t)=\bR),\quad t>0\nonumber\\
&&E(1|\bX(t_{1})=\bR_{1},\bX(t_{2})=\bR_{2})=\frac{\exp(-\frac{|\bR_{1}-\bR_{2}|^{2}}{2(t_{2}-t_{1})})}
{(2\pi(t_{2}-t_{1}))^{3/2}}, \quad t_{2}>t_{1}
\label{A.10}
\eeq
The fact that Brownian motion is an homogeneous process implies the following symmetry relations under
time and space translation, and space inversion:
\beq
E(F|\bX(t_{1})=\bR_{1},\bX(t_{2})=\bR_{2})&=&E(F_{\tau}|\bX(t_{1}+\tau)=\bR_{1},\bX(t_{2}+\tau)=\bR_{2})\nonumber\\
&=&E(F_{\bR}|\bX(t_{1})=\bR_{1}+\bR,\bX(t_{2})=\bR_{2}+\bR)\nonumber\\
&=&E(F_{-}|\bX(t_{1})=-\bR_{1},\bX(t_{2})=-\bR_{2})
\label{A.11}
\eeq
with $F_{\tau}(\bX(\cdot))=F(\bX(\cdot +\tau))$, $F_{\bR}(\bX(\cdot))=F(\bX(\cdot)+\bR)$ and $F_{-}(\bX(\cdot))
=F(-\bX(\cdot))$. 
Consider now the functional
\be
F_{q_{1},q_{2}}(\bX)=\exp\left[-\frac{\beta}{2}\int_{q_{1}}^{q_{2}}ds\int_{q_{1}}^{q_{2}}dt\tilde{\delta}(s-t)
(1-\delta_{[s_{1}],[s_{2}]}^{Kr})V(\bX(s)-\bX(t))\right]
\label{A.12}
\ee
where the index $(q_{1},q_{2})$ means that the functional depends on the path only when $t$ is in the interval $[q_{1}, q_{2}]$.
In the present notation the normalized Brownian bridge average of a functional $F$ reads
\be
\int D_{q}(\bX)F(\bX)=(2\pi q)^{3/2}E(F|\bX(0)=0,\bX(q)=0)
\label{A.12a}
\ee
and therefore the vertex function $\kappa(q)$ (\ref{5.2})
for a q-particle loop $\bX$ is 
\be
\kappa(q)=(2\pi q)^{3/2}E(F_{0,q}|\bX(0)=0,\bX(q)=0)
\label{A.13}
\ee
In $F_{0,q}(\bX)$ we suppress the interaction between the sets of particles $1,\ldots,q_{1}$ and
$q_{1}+1,\ldots,q$.  Since $V$ is positive this leads to the inequality
\be
F_{0,q}(\bX)\leq F_{0,q_{1}}(\bX)F_{q_{1},q}(\bX)
\label{A.14}
\ee
implying in (\ref{A.13}) in view of the Markov property of the Brownian motion
\beq
\kappa(q)&\leq & (2\pi q)^{3/2}\int d\bR E(F_{0,q_{1}}|\bX(0)=0,\bX(q_{1})=\bR) E(F_{q_{1},q}|\bX(q_{1})=\bR,\bX(q)=0)\nonumber\\
&=&(2\pi q)^{3/2}\int d\bR E(F_{0,q_{1}}|\bX(0)=0,\bX(q_{1})=\bR) E(F_{0,q_{2}}|\bX(0)=0,\bX(q_{2})=\bR)\nonumber\\
\label{A.15}
\eeq
with $q=q_{1}+q_{2}$. The second line follows from the symmetry relations (\ref{A.11}) and the fact that the potential is invariant
under space translations and inversion.
Considering now the Wiener expectation of $F_{0,q}(\bX)$ we establish in the same way
\be
\langle F_{0,q_{1}+q_{2}}\rangle_{W}\leq \langle F_{0,q_{1}}\rangle_{W}\langle F_{0,q_{2}}\rangle_{W}
\label{A.16}
\ee
again as a consequence of the Markov property of the Brownian process and the invariance (\ref{A.11}).
We exploit the inequalities (\ref{A.15}) and (\ref{A.16}) as follows. We first use (\ref{A.15}) to relax 
the constraint of closed path, noting from (\ref{A.10}) that  
$ E(F_{0,q_{2}}|\bX(0)=0,\bX(q_{2})=\bR)\leq (2\pi q_{2})^{-3/2}$ and from (\ref{A.15})
\be
\kappa(q)\leq \(1+\frac{q_{1}}{q_{2}}\right)^{3/2}\langle F_{0,q_{1}}\rangle_{W},\quad q=q_{1}+q_{2},\,\,q_{2}\geq 1
\label{A.17} 
\ee
Then we set $q_{1}=nq_{0},\;q_{2}=n,\,q=n(q_{0}+1)$ for some fixed integer $q_{0}\geq 1$. From (\ref{A.17}) and the iteration
of (\ref{A.16})
\be
\kappa(n(q_{0}+1))\leq (1+q_{0})^{3/2}(\langle F_{0,q_{0}}\rangle_{W})^{n}
\label{A.18}
\ee
If one sets
\be
\nu_{-}= -\frac{1}{\beta (q_{0}+1)}\ln \langle F_{0,q_{0}}\rangle_{W}\;>0,\quad r=q_{0}+1
\label{A.19}
\ee
one obtains the result of the lemma 2.

\vspace{3mm}

\noindent {\bf \large Appendix 3: The Bose gas in an external potential}

\vspace{3mm}

Motivated by the experimental situation of Bosonic atoms in traps, 
we briefly show in this Appendix how the formalism works in
presence of an external field
(more details can be found in [25]).
If a one-body external potential $V^{\rm ext}(\br)$ is introduced, the formula for the grand canonical partition
function $\Xi$ is still given by (\ref{2.100}) with the only change that  the loop activity $z(\cl)$ is replaced
by
\beq
\tilde{z}(\cl)&=&z(\cl)e^{-\beta V^{\rm ext}(\cl) }\nonumber\\
V^{\rm ext}(\cl)&=& \int_{0}^{q}dsV^{\rm ext}(\bR +\lambda \bX(s))
\label{A.20} 
\eeq
Since the the density is non uniform, it is appropriate to consider here the average total particle number $\langle N\rangle$
\beq
\langle N\rangle &=&\beta^{-1}\frac{\partial}{\partial \mu}\ln \Xi\nonumber\\
&=& \sum_{n=1}^{\infty}\frac{1}{n!}\int d\cl_{1}\cdots d\cl_{n}\(\sum_{\ell =1}^{n}q_{\ell}\right)\prod_{k=1}^{n}\tilde{z}(\cl_{k})
u(\cl_{1},\ldots,\cl_{n})\nonumber\\
&=&\sum_{n=1}^{\infty}\frac{1}{(n-1)!}\int d\cl_{1}\cdots d\cl_{n}q_{1}\prod_{k=1}^{n}\tilde{z}(\cl_{k})
u(\cl_{1},\ldots,\cl_{n})
\label{A.21}
\eeq
where we have introduced the Mayer expansion of $\ln\Xi$ and
$u(\cl_{1},\ldots,\cl_{n})$ is the Ursell function (\ref{2.15a}).

The external potential will be chosen positive and confining with sufficiently fast growth at infinity
so that $\int d\br \exp(-\beta V^{\rm ext}(\br))<\infty$. In order to allow for a well defined infinite particle number limit
it is necessary to scale the external potential as 
\be
V^{\rm ext}_{\omega}(\br)=V^{\rm ext}(\omega\br)
\label{A.22}
\ee
with $V^{\rm ext}(\br)$ a fixed positive confining potential.
As the scaling parameter $\omega$ tends to $0$, $V^{\rm ext}_{\omega}(\br)$ becomes less confining and  the average particle number
$\langle N\rangle(\beta,\mu,\omega)$ diverges. The proper quantity that remains finite in the limit is the product
$\omega^{3}\langle N\rangle(\beta,\mu,\omega)$. The experimental situation for atoms in traps appears to be well described in this
asymptotic regime [1]. We have indeed 

\vspace{3mm}

{\bf Proposition 5}

\noindent {\it In the range of convergence of the Mayer series for the uniform Bose gas,
$\lim_{\omega\to 0} \omega^{3}\langle N\rangle(\beta,\mu,\omega)\equiv {\cal N}(\beta,\mu)$ exists and
\be
{\cal N}(\beta,\mu) =\int d\br \rho(\beta,\mu-V^{\rm ext}(\br))
\label{A.23}
\ee
with $\rho(\beta,\mu)$ the density of the uniform Bose gas.}

\vspace{3mm}

\noindent In the scaling limit defined by (\ref{A.22}), the gas is locally uniform at point $\br$ with
a space dependent chemical potential $\mu(\br)\equiv \mu-V^{\rm ext}(\br)$.

\vspace{3mm}

\noindent We write $V^{\rm ext}_{\omega}(\cl)=V^{\rm ext}(\omega\bR,q,\omega\bX)$ and
$u(\cl_{1},\ldots,\cl_{n})=u(\bR_{1},q_{1},\bX_{1},\ldots,\bR_{n},q_{n},\bX_{n})$.
The spatial part of the integral in the n$^{\rm th}$ order term of the series (\ref{A.21})
reads (with the additional factor $\omega^{3}$)
$$
\omega^{3}\int d\bR_{1}d\bR_{2}\cdots d\bR_{n}\prod_{k=1}^{n}e^{-\beta V^{\rm
ext}(\omega\bR_{k},q_{k},\omega\bX_{k})}
u(\bR_{1},q_{1},\bX_{1},\bR_{2},q_{2},\bX_{2},\ldots,\bR_{n},q_{n},\bX_{n})\nonumber\\
$$
\beq
=\int
d\br e^{-\beta V^{\rm ext}(\br,q_{1},\omega\bX_{1})} &\times&\int d\bR_{2}\cdots d\bR_{n}\prod_{k=2}^{n}e^{-\beta V^{\rm
ext}(\br+\omega\bR_{k},q_{k},\omega\bX_{k})}\nonumber\\ 
&\times &
u(0,q_{1},\bX_{1},\bR_{2},q_{2},\bX_{2},\ldots,\bR_{n},q_{n},\bX_{n}) 
\label{A.24}
\eeq
where one has made the sucessive change of variables $\bR_{k}\rightarrow \bR_{k}+\bR_{1},\;k=2,\ldots,n$, then $\omega\bR_{1}=\br$
and used  that the Ursell function is translation invariant with respect to its spatial variables.
Since $\lim_{\omega\to 0}V^{\rm ext}(\br+\omega\bR_{k},q_{k},\omega\bX_{k})= q_{k}V^{\rm ext}(\br)$ and the Ursell
function is jointly integrable on $\bR_{2},\dots, \bR_{k}$
the expression (\ref{A.24}) tends by dominated convergence to
\be
\int d\br \left[\prod_{k=1}^{n}e^{-q_{k}V^{\rm ext}(\br)}\right]\left[\int d\bR_{2}\cdots d\bR_{n} 
u(0,\bX_{1},q_{1},\bR_{2},\bX_{2},q_{2},\ldots,\bR_{n},\bX_{n},q_{n})\right]
\label{A.25}
\ee
The first bracket, when combined with the loop activity $z(\cl)$ (\ref{2.11}), simply yields the shifted local chemical potential
$\mu(\br)=\mu-V^{\rm ext}(\br)$ whereas the second bracket is the spatial part of the integral of the Ursell function
occuring in the loop density series (\ref{2.14a}) of the uniform gas evaluated at point $\bR_{1}=0$.
When this is introduced in the series (\ref{A.21}) for $\omega^{3}\langle N\rangle(\beta,\mu,\omega)$  
and the integrations on the internal degrees of freedom of the loops are taken into account (using also (\ref{2.16})),
one obtains the result (\ref{A.23}) in the limit $\omega \to 0$.

To show convergence, one introduces the expresssion (\ref{A.24}) in the series (\ref{A.21}) and majorize factors 
$e^{-\beta V^{\rm ext}(\br+\omega\bR_{k},q_{k},\omega\bX_{k})}$  by $1$, $\;k=2,\ldots,n$ :
\beq
\omega^{3}\langle N\rangle(\beta,\mu,\omega)&\leq& \sum_{n=1}^{\infty}\frac{1}{(n-1)!}
\sum_{q_{1}=1}^{\infty}q_{1}\int D_{q_{1}}(\bX_{1}) \int d\br  e^{-\beta V^{\rm ext}(\br,q_{1},\omega\bX_{1})}\nonumber\\
&\times&\prod_{k=1}^{n}z(\cl_{k})\int d\cl_{2}\cdots d\cl_{n}u(\cl_{1},\cl_{2},\ldots,\cl_{n})\nonumber\\
&=&\sum_{q_{1}}^{\infty}q_{1} \int D_{q_{1}}(\bX_{1}) 
\int d\br e^{-\beta V^{\rm ext}(\br,q_{1},\omega\bX_{1})}\rho_{{\rm loop}}(\cl_{1})
\label{A.26}
\eeq
where $\rho_{{\rm loop}}(\cl_{1})$ is the Mayer series (\ref{2.14a}) for the uniform system evaluted at $\cl_{1}=(0,
q_{1},\bX_{1})$. Using again the positivity of $V^{\rm ext}(\br)$ and Jensen inequality we have
\beq
&&\int D_{q_{1}}(\bX_{1})\int d\br  e^{-\beta V^{\rm ext}(\br,q_{1},\omega\bX_{1})}\nonumber\\
&\leq &\int D_{q_{1}}(\bX_{1}) \int d\br 
\exp(-\beta\int_{0}^{1}dsV_{\rm ext}(\br +\omega\lambda \bX(s))\nonumber\\
&\leq & \int D_{q_{1}}(\bX_{1})\int_{0}^{1}ds\int d\br\exp(-\beta V_{\rm ext}(\br +\omega\lambda \bX(s))
=\int d\br e^{-\beta V^{\rm ext}(\br))}
\label{A.27}
\eeq
The convergence of  Mayer series (\ref{2.14a}) is established in Proposition 2 with bounds that are independant
of the shape $\bX_{1}$ of the loop $\cl_{1}$. Combining this previous analysis with (\ref{A.27}), one gets eventually
\be
\omega^{3}\langle N\rangle(\beta,\mu,\omega)\leq  \bar{\rho}(\mu)\int d\br e^{-\beta V^{\rm ext}(\br)}
\ee
with $ \bar{\rho}(\mu)$ defined in Proposition 2. Since estimates are uniform with respect to $\omega$, the existence of the limit
(\ref{A.23}) follows again by dominated convergence.

\vspace{3mm}

Moreover if the two-body potential is scaled according to (\ref{0.5a}), one obtains immediately from Proposition 3
the mean field limit for the trapped Bose gas
\be
{\cal N}_{{\rm mf}}(\beta,\mu)\equiv \lim_{\gamma\to 0}{\cal N}(\beta,\mu,\gamma)= \int d\br \rho_{{\rm mf}}(\beta,\mu -V^{\rm
ext}(\br))
\label{A.28}
\ee
This result is established by other methods in [9] in a sligtly different situation: the external potential considered in [9] has
support in a box of volume $L^{3}$ and is scaled according to the size of the box (namely $\omega=\frac{1}{L}$ in (\ref{A.22})).

The thermodynamics of the mean field trapped gas can be studied in detail from (\ref{A.28}). If $\mu\leq \mu_{c}=a\rho_{0,c}$ (the
critical chemical potential of the homogeneous mean field gas), then $\mu(\br)=\mu-V^{\rm ext}(\br)\leq \mu_{c}$ for all $\br$,
so condensation does not occur anywhere. If $\mu>\mu_{c}$ there is a local condensate of density $\left[\mu(\br)/a
-\rho_{0,c}\right]$ at all points $\br$ such that $\mu(\br)>\mu_{c}$, namely in the region of space $\Delta=\{\br\;| V^{\rm
ext}(\br)<\mu-\mu_{c}\}$. Then the total amount of condensate is
\be
{\cal N}_{{\rm cond}}(\beta,\mu)=\int_{\Delta}d\br\left[\mu(\br)/a -\rho_{0,c}\right]  
\label{A.29} 
\ee
Obviously ${\cal N}_{{\rm cond}}(\beta,\mu)$ vanishes at $\mu=\mu_{c}$ so the critical temperature $T_{c}(\cal N)$ 
as function of the total particle number is defined as the solution of (\ref{A.28}) when $\mu=\mu_{c}$ and
${\cal N}$ is fixed, namely 
\be
{\cal N}_{{\rm mf}}(\beta_{c},\mu(\beta_{c}))= {\cal N},\quad \mu(\beta)=a\rho_{0,c}(\beta)
\label{A.30}
\ee
As an example, for an harmonic potential $V^{\rm ext}(\br)=b|\br|^{2}/2$, one finds a critical behaviour
of the condensate fraction ${\cal N}_{{\rm cond}}(\beta, {\cal N})/{\cal N}$ 
of the form (keeping now the particle number ${\cal N}={\cal N}_{{\rm mf}}(\beta,\mu)$ fixed) [25]
\be
\frac{{\cal N}_{{\rm cond}}(\beta, {\cal N})}{{\cal N}} \sim \(1-\frac{T}{T_{c}}\right)^{5/2},\quad T\to T_{c}
\ee
This is to be contrasted with the behaviours of the same fraction for the homogeneous free or mean field gas  
($\sim 1-\(\frac{T}{T_{0,c}}\right)^{3/2}$)  and for the free gas in an harmonic potential ($\sim 1-\(\frac{T}{T_{{\rm
trap},c}}\right)^{3}$) when $T$ approaches the corresponding critical temperatures. 

Clearly, when corrections to the mean field density of the homogeneous gas are known, they can be implemented in equation
(\ref{A.23}) and thus will also give interesting informations on imperfect trapped gases.

\end{document}